# QUERY SENSITIVE COMPARATIVE SUMMARIZATION OF SEARCH RESULTS USING CONCEPT BASED SEGMENTATION

Chitra.P[1], Baskaran.R[2], Sarukesi.K[3]

[1]Dept. of Information Technology, RMK Engineering College, Tamilnadu, India
[2]Dept. of Comp. Sci. & Engg. Anna University, Chennai, Tamilnadu, India
[3]Hindustan University, Chennai, Tamilnadu, India

[1]pce.it@rmkec.ac.in, [2]baaski@cs.annauniv.edu, [3]profsaru@gmail.com

## ABSTRACT

*Query sensitive summarization aims at providing the users with the summary of the contents of single or multiple web pages based on the search query. This paper proposes a novel idea of generating a comparative summary from a set of URLs from the search result. User selects a set of web page links from the search result produced by search engine. Comparative summary of these selected web sites is generated. This method makes use of HTML DOM tree structure of these web pages. HTML documents are segmented into set of concept blocks. Sentence score of each concept block is computed with respect to the query and feature keywords. The important sentences from the concept blocks of different web pages are extracted to compose the comparative summary on the fly. This system reduces the time and effort required for the user to browse various web sites to compare the information. The comparative summary of the contents would help the users in quick decision making.*

## 1. INTRODUCTION

The WWW grows rapidly and caters to a diversified levels and categories of users. Web search engines helps in locating information content and normally provide thousands of results for a query. Users still have to spend lot of time to scan through the contents of this result set to locate the required information. It is not feasible for the user to open each link in the result set to find out its relevance. The performance improvement of the search engines has become the most important research area to satisfy the needs and expectations of diversified target users.

A small summary generated from the content of web page would be helpful for the users to get an instant feel about the content without going through the entire content. People can have a concise overview in short time. This can greatly enhance the retrieval efficiency. Automatic summary aims to extract some important sentences from original documents to represent the content of the article.

                                           

Automatic summary produced by current search engines contains first few sentences of the web page or the set of sentences containing the query key words. Using this information, users have to decide which of the listed documents in the search result will be most likely to satisfy their information need.

This paper is an extension of our work mentioned in [1], proposes a summarization technique, which extracts query relevant important sentences from a set of selected web pages to generate a comparative summary which would be beneficial for the users to make informed decisions.

The remainder of this paper is organized as follows: Section 2 provides the motivating examples for this work, Section 3 discusses about the related research works that have been done in this field and Section 4 describes concept based segmentation process guided by the webpage's DOM tree structure. In section 5, we present the framework for the selection based comparative summarization system. Section 6 compares this system with few other systems, Section 7 discusses about experimentation results and performance measures and in section 8, the paper is concluded with a view on further improvement to this system.

## 2. MOTIVATING EXAMPLE

People normally collect all related material and information before they make a decision about some product or service before they go for it. For example, parents might be interested in collecting details like placement, infrastructure, faculty details, etc related to various Engineering Colleges, before they go for admission. To accomplish this, the user collects placement and other required details of various Engineering Colleges using search engines like Google, prepares a comparative statement manually to find out the best option for admission.

The proposed system generates the comparative summary from the set of URLs selected by user from the search result based on the specified feature set. The comparative summary contains the text relevant to placement and training, infrastructure details, result details and fee structures from the selected URLs. This would definitely be helpful to get instant comparative statement.

Another example could be the comparison between the services offered by various Banks. Set of Banks can be selected from the list of Bank web sites and comparative summary based on feature keywords like home loan, term deposit, etc would be helpful for users to make quick decisions about their investment.

## 3. RELATED WORKS

Summarization in general can be categorized into two types as extraction based and abstraction based methods. Extractive summary is created by extracting important sentences from the actual content, based on some statistical measures like TFxIDF, SimWithFirst[2],etc. Abstractive summary is created by rewriting sentences on understanding the entire content of the original article by applying NLP techniques. The later technique is more computationally intensive for large data-sets.

Concept based automatic summarization directly extracts the sentences, which are related to the main concept of the original document while the query sensitive summarization extracts sentences according to user queries, so as to fit the interests of users. In multi-document summarization the sentences are selected across different documents by considering the concept and diversity of contents in all documents.

Query based summarization system to create a new composed page containing all the query key words was proposed in[3]. Composed page was created by extracting and stitching together the relevant pieces from a particular URL in search result and all its linked documents but not other relevant documents of user's interest.

Segmented topic blocks from HTML DOM tree were utilized to generate summary in [4] by applying a statistical method similar to TFxIDF to measure the importance of sentences and MMR to reduce redundancy. This system focused on the summary of only single document. SimWithFirst (Similarity With First Sentence) and MEAD (Combination of Centroid, Position, and Length Features) called CPSL features were used for both single and multi document text summarization in[5]. Both these techniques show better performance for short document summarization but not suitable for large ones.

Document Graph structure of sentences was used in[6] for text summarization. Similarity scores between the query and each sentence in the graph are computed. Document graph construction is an overhead for the summarization process.

Balanced hierarchical structure[6] was utilized to organize the news documents based on event topics to generate event based summarization. This method focused on news and event summarization.

This research work focuses on the novel idea of generating the aggregation of document summaries. This document summarization makes use of concept based segmentation of DOM tree structure of web pages. This comparative summary is composed of the query sensitive important sentences extracted from concept blocks of different web pages which would be helpful for decision making.

## 4. CONCEPT BASED SEGMENTATION

In general, web page summarization derives from text summarization techniques, while it is a great challenge to summarize Web pages automatically and effectively. Because Web pages differ from traditional text documents in both structure and content. Web pages often have diverse contents such as bullets, images and links.

Web documents may contain diversified subjects and information content. Normally, the contents of same subjects will be grouped under the same tag. This system utilizes the Document Object Model (DOM) to analyze the content of the web page. The leaf nodes of DOM tree contain the actual content and the parent nodes generally contain higher level topics or section headings.

### 4.1 Concept Based Segmentation Process

Fig. 1 depicts the concept based segmentation using DOM tree structure. The rectangular nodes represent the HTML tags or the higher level topics and the circular nodes represent the information content within the tag. These circular nodes from left to right constitute a coherent semantic string of the content[4].

The DOM trees of the user selected URLs are processed to generate the summary. Leaf nodes are considered as micro blocks which are the basic building blocks of the summary. Adjacent micro blocks of the same parent tag are merged to form the topic blocks.

Each sentence in the topic block is labeled automatically based on the PropBank notations [7][8]. The information about who is doing what to whom clarifies the contribution of each term in a sentence to the meaning of the main topic of that sentence[9].This concept-based mining model captures the semantic structure of each term within a sentence rather than the frequency of the term alone. This similarity measure outperforms other similarity measures that are based on term analysis models[10].

These sentences are labeled by a semantic role labeler that determines the words which contribute more to the sentence semantics associated with their semantic roles in the sentence.

The semantic role labeler identifies the verb argument structures of each sentence in the topic block. The number of generated labeled verb argument structures is entirely dependent on the amount of information in the sentence. The sentence that has many labeled verb argument structures includes many verbs associated with their arguments.

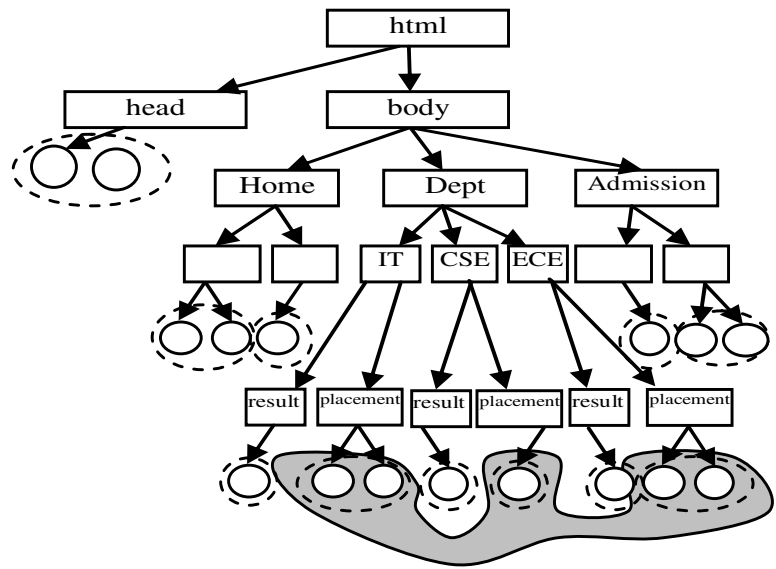

**Fig 1. Concept based segmentation**

The words contributing more to the meaning of the sentence will occur more number of times in the verb argument structure of the sentence. Hence these words will have comparatively higher frequency. ASSERT software which is a publicly distributed semantic labeling tool, is used for this purpose. Each word that has a semantic role in the sentence, is called a concept[4][9]. Concepts can be either words or phrases and are totally dependent on the semantic structure of the sentence. List of concept words and their respective frequency of occurrences for these topic blocks are identified.

Concept based similarity between the topic blocks are measured using the concept lists to identify the similar topic blocks. Topic blocks having similarity above the threshold value α(0.6), are combined to form the concept block. Topic blocks having content about the same concept (for example placement and training in a college web site) will be similar to each other. Topic blocks containing information about similar concept word (placement) are merged to form a concept block (placement block having placement details of all departments in the college).

The concept block formation could be done offline for all web documents in the repository. The concept block id, conceptual terms, frequency and list of sequence numbers of sentences of each of these concept blocks are stored in the offline database which would be required for processing at run time. These concept blocks contain related information content scattered throughout the document. The set of sentences in each of these concept blocks are actually present is different parts of the document.

Since DOM nodes are processed, the time taken for processing is less when compared to other vector based and document graph[10] based models. The processing time required to build the document graph is avoided in this approach.

## 4.2 Concept Based Segmentation Algorithm

The conceptual term frequency is an important factor in calculating the concept-based similarity measure between topic blocks. The more frequent the concept appears in the verb argument structures[4][9] of a sentence, the more conceptually similar the topic blocks are. Concept based segmentation algorithm is described below:

Input : Web document $d_i$.
Output : Set of concept blocks{$Cb_1$...$Cb_n$} of di,

    Concept list of $d_i$, L={$C_1$...$C_m$}
    Concept list of topic block $tb_i$, $C_{tbi}$ ={$c_{k1}$,..$c_{km}$}, k=1..n

**Step1:** Mark all leaf nodes as individual micro blocks in the DOM tree.
**Step2:** Extend the border of the micro block to include all leaf nodes of the same tag to form a topic block so as to have a set of topic blocks TB={$tb_1$, $tb_2$, …$tb_n$}, TB⊂di.
**Step3:** Build concept list for all topic blocks TB ={$tb_1$, $tb_2$, …$tb_n$}.
Topic block $tb_i$ is a set of sentences, tbi={$s_{i1}$, $s_{i2}$, ..$s_{in1}$}, $s_i$⊂ $tb_i$
Sentence $s_i$ is a string of concepts, $s_i$={$c_{i1}$,$c_{i2}$ . . . $c_{im}$}, $c_i$ ⊂ $s_i$, if $c_i$ is a substring of $s_i$
Concept $c_i$ is a string of words, $c_i$ = {$w_{i1}$,$w_{i2}$, . . .$w_{ik}$ },
   where k : number of words in concept $c_i$.
   m: number of concepts generated from the verb argument structures in sentences
   n1: total number of sentences in $tb_i$
  3.1: $C_{tbi}$, L are empty lists.
  3.2: Build concept list of each sentence in $tb_i$
      3.2.1: si is a new sentence in tbi
      3.2.2: Build concepts list Ci from si, Ci ={c1,c2,..cm}
      3.2.3: Update concept list $C_{tbi}$ of topic block tbi and L of document $d_i$
         3.2.3.1: for each concept ci ⊂ Ci do
         3.2.3.2: for each cj⊂ L, do
         3.2.3.3: if (ci == cj) then
         3.2.3.4:   add freq(ci, si) to ctfi of ci
                  // freq(ci, si) returns the frequency of ci in the verb argument
                  // structures of si, added to conceptual term frequency of ci
         3.2.3.5: else add new concept to $C_{tbi}$, L // added to both L and $C_{tbi}$
         3.2.3.6: end if
         3.2.3.7: end for
         3.2.3.8: end for
  3.3: Output the concepts list $C_{tbi}$
  3.4: Output the concepts list L of document di.

**Step4:** The concept based similarity between topic blocks are measured by (1) .

$$Sim(tb_1,tb_2) = 1 - \left| \sum_{i1 \subset C_{tb1} \cap C_{tb2}} ctf_{i1} \times ctfweight_{1} - \sum_{i2 \subset C_{tb1} \cap C_{tb2}} ctf_{i2} \times ctfweight_{2} \right|$$

(1)

Where,

$ctf_{i1}$, $ctf_{i2}$ : Frequency of concept $c_i$ in $tb_1$, $tb_2$

i1,i2 : set of common conceptual terms between $tb_1$, $tb_2$

$ctfweight_{i1}$,$ctfweight_{i2}$ : Weight of concept $c_i$ with respect to topic blocks $tb_1$, $tb_2$ normalized by the frequency vectors of $tb_1$, $tb_2$ calculated as in (2)

$$ctfweight_i = \frac{ctf_i}{\sqrt{\sum_{k=1}^{m}(ctf_k)^2}}$$ (2)

$ctfweight_i$ represents the importance of the concept ci with respect to the concept vector of tbi. The concepts contributing more to the meaning of the sentences in the topic block occurs more times in the verb argument structure of the sentence and in turn gets more weightage.

**Step5:** Merge the topic blocks having concept based similarity measure above the predefined threshold α.

Concept block $Cb_k$={set of topic blocks $tb_i$}|

∀tbi, tbj∈$Cb_k$, sim(tbi, tbj)> α, $tb_i$⊂TB, $tb_j$⊂TB, k=1..n

**Step6:** Output Concept blocks $Cb_1$,$Cb_2$,..$Cb_n$

Similarity between topic blocks is measured by considering the commonly occurring concepts in both topic blocks, tb1 and tb2. Frequency of these common terms and their topic block based weightage are used to measure the similarity score and is normalized to the range 0 to 1.

The concept blocks of each URLs selected by the user are identified. Concept blocks of all URLs in the repository can be identified during preprocessing stage itself. This will reduce the computation complexity at run time.

The next section describes about generating the comparative summary on the fly at run time using these concept blocks of the web document.

## 5. COMPARATIVE SUMMARY GENERATION

The architecture of the comparative summarization system is given in Fig.2. User enters the generic query string (eg. Engineering College) through the search engine query interface. Search engine identifies the relevant pages and present the search result in rank order. Then the specific feature keywords based on which comparative summary is to be generated and the set of URLs are obtained from the user.

The selected HTML files of the URLs are cleaned by removing unwanted HTML tags (like META tag, ALIGN tag, etc.) which do not contribute much for further processing. Concept blocks of these URLs which were already formed during preprocessing are utilized to generate the summary.

The relevance of the concept blocks Cbi to the feature keywords f is measured by means of similarity between the feature keyword string[4][9] and the concept list of each of the concept blocks Cbi..

$$sim_f(f, Cb_i) = \frac{\sum_{ti \in f \cap Cbi} Ctf_{ti}}{\sqrt{\sum_{ti \in Cbi} Ctf_{ti}^2}}$$ (3)

Where,

f: set of feature keywords,f={f1,f2,..fn}

$Cb_i$: Concept block i, for which the list of concept terms and their frequency were already identified.

$Sim_f$ (f, $Cb_i$) : Similarity between feature keyword string f and Concept block $Cb_i$

t          : set of common terms between f and Concept list of $Cb_i$

$Ctf_{it}$     : frequency of term t in $Cb_i$

$Sim_f$(f,Cbi) is measured using the conceptual term frequency of the matching concepts of these concept blocks and is normalized by the concept frequency vector of the concept block. The concept block having maximum number of matching concept terms will get high score. The range of $Sim_f$(f,Cbi)  value lies between 0 and 1, and the similarity increases as this value increases. Synonyms of the feature words and concept terms were taken into consideration for processing. The concept block with maximum similarity is considered as the superset of the summary to be generated.

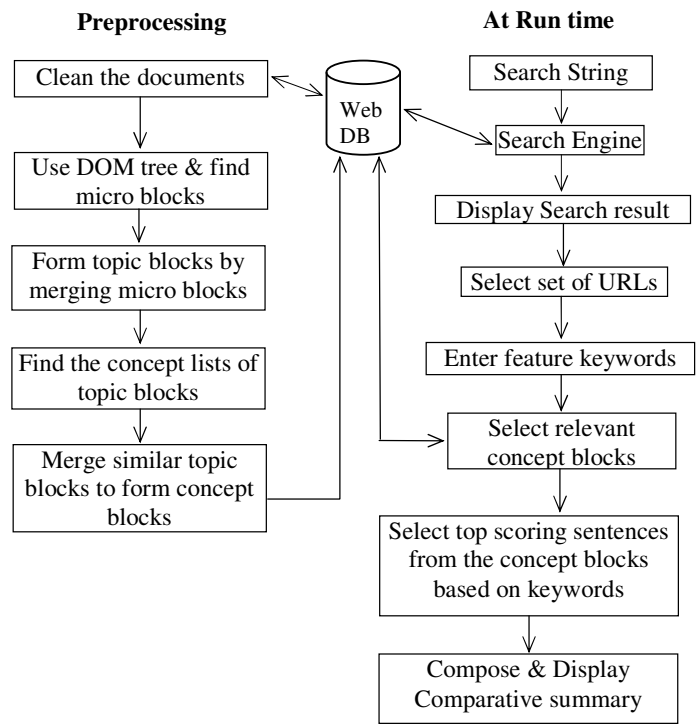

Fig. 2. Framework of comparative summarization system

The significance of each sentence in this concept block with respect to the query string is measured. The sentences are considered in the descending order of their score. According to the ratio of summarization required or the number of sentences required, the sentences are extracted from these concept blocks to compose the HTML page for comparative summary.

## 5.1   Sentence Weight Calculation

Content of these concept blocks are ranked with respect to the query string and the feature keywords, using (4). Sentence weight calculation considers[12] the number of occurrences of query string, feature keywords and their distance and frequency, location of the sentence, tag in which the text appears in the document, uppercase words,etc.

$$W(S_i) = \frac{\sum_{i=1}^{n} W(k_i) + e^{-\gamma(D-1)} + \alpha W_{tag} + \beta W_l}{Len(S_i)} \qquad (4)$$

Where,

$W(ki)$ :Weight of certain feature word or query word $ki$ in sentence $si$
$Len(si)$ :number of words contained in the sentence.
$W_{tag}$ :weight determined by the tag to which the words of sentence belong
     BOLD, UNDERLINE, ITALICS,  CAPTION, PARAGRAPH TITLE : 3
     COLOR CHANGE :2
$W_l$ :weight determined by the location of  the Sentence with respect to the parent
     node, set to 1 to the left most node and 0.5 to the right most node of a parent
$D$ :average distance between feature words measured by counting the number
     of other words in between the feature words

$\alpha$ , $\beta$ and $\gamma$ :adjusting parameters.

Table 1. Comparison between Search Engine Snippets, Mead and
Comparative Summarizer

| Parameters | Search Engine Snippets | MEAD Summarizer | Comparative summarizer |
|---|---|---|---|
| Method of summary generation | Extraction | Extraction | Extraction |
| Techniques used | Occurrence of query string | Lex rank, Centroid Position | DOM tree, Concept based segmentation |
| Document type | Web page | Text Documents | Web Pages |
| Single/multi document | Single | Multi Document | Multi Document |
| User control | No Control | Less Control | More Control |
| Satisfaction Index | Low | Medium | Medium to high |
| Run time over head | Less | Comparatively high | Comparatively high |
| Usage | Gives a clue about the relevance of the document | Short summary is generated | Comparative Summary is generated which is useful for decision making |
| Retrieval Efficiency | Need further | Reduces time | Reduces time and |

| | browsing and scanning | taken for scanning entire set of documents to understand the core concept | effort taken for browsing and scanning various web pages to extract the gist of it |
|---|---|---|---|

 The sentences having frequent occurrences of the feature keywords and enclosed in special tags are given preference. Location based weight is assigned according to the location of the sentence within its immediate parent node. The top scoring sentences are extracted and arranged based on the hierarchical structure of the individual documents. The title or first sentence of the immediate parent of the extracted sentence, is chosen as subtitles for a set of leaf node contents. (for example, IT, CSE, ECE in our example). Hence the resulting summary will contain the SECTION-wise summary sentences of a set of URLs chosen by the user for immediate comparison. This is applicable to various decision making situations which require analysis of various parameters from various sources.

## 6. COMPARISON WITH EXISTING SYSTEMS

Various feature of the proposed system and the search snippets and a bench mark text summarization system are compare and is presented in Table 1. As there is no bench mark web document summarizer is available the techniques used in MEAD[14] is compared with the current system. The snippets are the set of sentences displayed by search engines as part of search results along with URLs that are extracted from the web page. These are the line in which the search string occurs in the web page.

As given in Table1, the snippets provide vague information which are not sufficient to guess the usefulness of the target page and requires a complete scan of the content to capture the required information.

MEAD provides a summary of set of text documents based on Lexrank and Centroid position score. This prepares the summary of the core concept of the documents which might be modified as per the query string.

This system makes use of concept based segmentation approach which give more importance to conceptual terms contributing more to the meaning of the sentences. Hence this systems performance is comparatively promising than the other summarization systems.

## 7. EXPERIMENTAL RESULTS AND DISCUSSION

The experimentation of this work was carried out with the real time dataset containing randomly collected 200 web documents from internet related to the educational institutions, algorithms, banking and household items.
Normally, the summarization systems are evaluated using intrinsic approach or extrinsic approach [6][11]. Intrinsic approach directly analyzes the quality of the automatic summary through comparing it with the abstract extracted by hand or generated from other different automatic system. Extrinsic approach is a task-oriented approach, which measures the abstract quality according to its contribution.

Intrinsic approach was utilized to conduct the experiment of this system. Users including one engineering student, three naïve users and one expert level user were involved in the experimentation process.

The evaluation feedback collected from the experimentation is listed in table 2. and is plotted Fig. 3.

Table 2. Evaluation feedback measure

| User# | Query | Feature Keywords | Feed back in 5 point scale |
|---|---|---|---|
| #1 | Engineering College, Chennai | Placement, Recruiters | 4 |
| #2 | Optimization Algorithms | Efficiency, Time Complexity | 3.5 |
| #3 | Theme Park, Chennai | Entry Fee, Games | 4.2 |
| #4 | Banks | Home loan rate, services | 3.5 |
| #5 | Washing Machine | Brands, cost, offers, warranty | 3.3 |

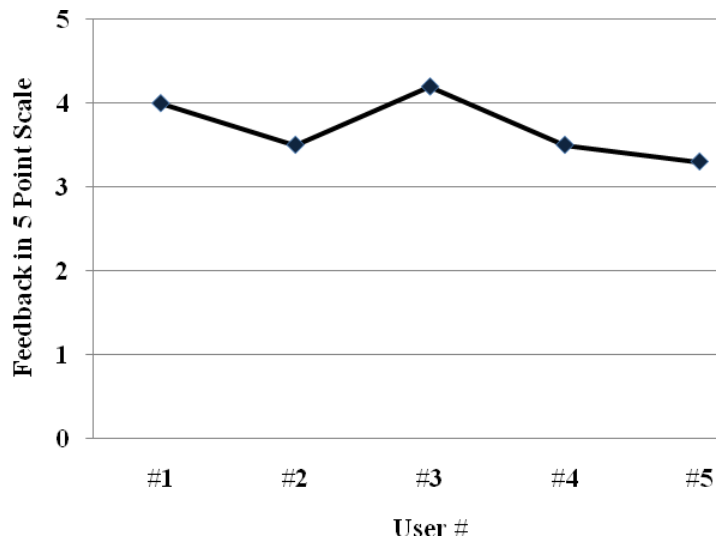

Fig. 3 User Feedback in 5 Point Scale

The query result user interface of the system is given in Fig. 4 through which the user selects the URLs and enter feature words for comparative summary generation.

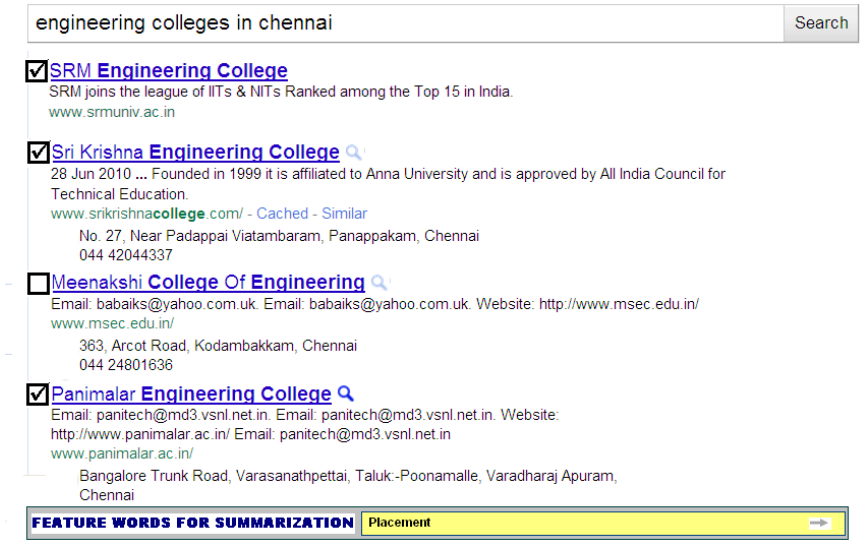

Fig. 4 Query result and URL selection interface

Summary extracted from these selected URLs are presented to the user as a comparative summary given in Fig 5. This system produces the comparison information required for real time decision making dynamically. The average user satisfaction index is 3.7.

Fig. 5 Comparative summary

Since DOM tree structure of the web documents are utilized to process the content and generate the summary the time complexity involved is very less when compared to systems making use of vector space model and document graph model.

## 8. CONCLUSION AND FUTURE WORK

This system focused on generating comparative summary from a set of URLs selected by the user. Concept based segmentation is used to identify the relevant block of content in the document and top scoring sentences are extracted, composed and displayed to the user. This summary would definitely be helpful for the users to get the immediate summary and for decision making.

The impact of usage of key words association, document graph model of documents on this system and advanced text clustering techniques for summary generation can be done as a future expansion to this system.

# Authors

**Ms. P. Chitra**, Associate Professor, RMK Engineering College, Chennai, India received B.E degree in Computer Science and Engineering from Bharathidasan University, Trichy, India, in 1992, and the M.E degree in Computer Science and Engineering from Anna University, Chennai, India, in 2006. Currently she is pursuing Ph.D programme in the Department of Information and Communication Engineering at Anna University – Chennai, India. She has presented 3 papers international conferences and authored 2 books. She has organized many Workshops and Seminars. She is a life membe r of ISTE (Indian Society for Technical Education) and IACSIT (International Association of Computer Science and Information Technology).

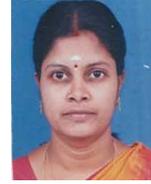

**Dr.R.Baskaran** received his B.Tech in Electrical and Electronics Engineering, Masters Degree in Computer Science and Engineering, Madras University and Doctor ate from Anna University in the years 2000,2001, and 2007 respectively. He is now working as a Assistant Profoessor in Department of Computer Science and Engineering, College of Engineering Guindy, Anna University Chennai. His present research includes Networks, Database and Image Processing. He has published 35+ papers in International, National Journals and Conferences. He is a life member of Institution of Electronics and Telecommunication Engineers (IETE), Indian Society for Technical Education (ISTE), Computer Society of India (CSI) and International Association for Engineers.

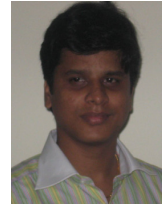

**Dr. K. Sarukesi**, Vice Chancellor, Hindustan University has a very distinguished career spanning over 40 years. He has a vast teaching experience in various universities in India and abroad. He was awarded a common wealth scholarship by the association of common wealth universities, London for doing Ph.D in UK. He completed his Ph.D from the University of Warwick – U.K in the year 1982. His area of specialization is Technological Information System. He worked as expert in various foreign universities. He has executed number of consultancy projects. He has been honored and awarded commendations for his work in the field of information technology by the government of Tamilnadu. He has published over 30 research papers in international conferences/journals and 40 National Conferences/journals.

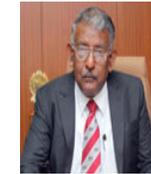